\documentclass[10pt]{article} 
\begin{document}
\large
%=========================================================================
~\\
\vskip 1.5 cm
\begin{flushleft}
{\bf\Huge
Cosmological contraction of the atomic space-time scale 
}
\end{flushleft}
\vskip 0.5 cm
%=========================================================================

\begin{flushright}
\begin{tabular}{p{5.5in}}
{\bf\Large S.S. Stepanov}\\
Dnepropetrovsk State University\\
e-mail: steps@tiv.dp.ua\\
~\\
\hline
\normalsize
\it
Quantum mechanical effects related to the recently developed
projective theory of relativity are considered. It is shown that
at cosmological time intervals the light velocity increases and
the atomic units of length and time are shrinking. A new derivation 
of the Hubble Law is presented.
\\
\hline
\end{tabular}
\end{flushright}

\vskip 1 cm

\section{Introduction}

The modern relativistic theory was initially based on the two postulates
put forward by Einstein~\cite{Einstein1905}, namely, the principles of 
relativity and constancy of light velocity.
However, shortly after the Einstein's theory was developed, it was shown
\cite{Pauli} that these postulates are not independent. Moreover, 
it turned out that the number of axioms underlying the relativity theory 
is actually less than that of the classical mechanics.

This property of the relativistic theory is shared by 
other fundamental physical theories too. The
principle of ``parametric incompleteness'' was formulated in 
Ref.~\cite{Steps1999}, which states
that fundamental physical constants like $c$ and $\hbar$ appear in the
theory as a result of reducing the number of 
axioms of classical mechanics. This principle leads to an heuristic
method for creating new physical theories:
{\it
If it is possible to reduce the number of axioms of a theory 
in such a way that all the functional relations (theorems)
within this theory can be derived, one obtains a new, more
general physical theory. New physical constants appear in this new 
theory as a consequence of reducing the information content of the
initial set of axioms.
}                                      
Classical mechanics is a complete theory and contains no fundamental
physical constants. Relativistic and quantum theories can be derived
from classical mechanics if one rejects some of its axioms, and so
they are parametrically incomplete theories depending on the 
constants $c$ and $\hbar$, the values of which cannot be inferred using only
those theories' intrinsic means.

Recently this method was applied to constructing a generalization of
the special theory of relativity \cite{Manida1999}-\cite{Steps1999:2}, 
which was named the projective theory of relativity \cite{Steps1999}.
Formally, this theory incorporates two parts, 
the generalized Lorenz transformations (between two inertial reference
frames) \cite{Manida1999},\cite{Steps1999}, and the generalized
translational transformations (between observers situated at different
places in the same inertial frame) \cite{Steps1999:2}. 
Together they compose a six-parametric group of linear fractional
transformations of the projective geometry.
In addition to the fundamental constant velocity $c$, the formulae of
the projective relativity contain a new constant of inverse-length 
dimension, $\kappa$.
It was denoted differently when first introduced, namely
$\kappa=-1/R$ in Ref. \cite{Manida1999}, and
$\kappa=-\lambda c$ in Ref. \cite{Steps1999}, \cite{Steps1999:2}.
We will show in the present paper that the constant $R$ ($\lambda$)
is in fact negative, which makes the new notation more natural. 

When $\kappa$ is small enough, in the limit $\kappa\to0$ the
convenient relativistic theory is restored, and corrections are
of any significance only for events that are far distant in time and/or
space from the observer. Therefore, we see that the projective relativity 
is a cosmological theory.

The pivotal property of the discussed theory is that the maximal
possible speed at a given point of space is a function of both the
temporal and spatial coordinates: 
\begin{equation}\label{speedlight}
\vec{C}(t,\vec{x})= c\frac{\vec{n}-\kappa \vec{x}}{1-\kappa c t},
\end{equation}
where $c$, $\kappa$ are fundamental constants, 
$\vec{n}$ is a unit vector along the flux of the electromagnetic wave
and $\vec{x}$ is the distance to the observer.
We will identify $\vec{C}(t,\vec{x})$ with the physical velocity of 
light. At present, in the vicinity of the observer
$\vec{C}(0,0)=c=299~792~458~ m/s$.
It is important that while being a function of coordinates
$C(t,\vec{x})$, the light speed is nevertheless invariant under the group of
transformations of the theory. Moreover, it does not change along the
trajectory of a light pulse~\cite{Steps1999:2}.
It is follows from Eq.~(\ref{speedlight}) that the speed of light
increases with time ($\kappa>0$). This fact could find its application
to recently developed models relying upon a time-dependent light 
velocity~\cite{Moffat}-\cite{Youm}.
In particular, the projective theory of relativity could reconcile the
principle of relativity and the speed of light varying with time 
\cite{Brandenberger1999}. 

As the light velocity is coordinate- and time-dependent, the frequency of
signals emitted by a fixed remote atomic source experiences a 
redshift, which is a growing function of the distance to the observer:
\begin{equation}\label{redshift}
\frac{\omega_0}{\omega}=1+z=\sqrt{\frac{1+\kappa R}{1-\kappa R}},
\end{equation}
where $\omega$ is the frequency of the signal at the moment of detection, 
$\omega_0$ - the frequency of an identical atom on Earth, 
$R$ - the distance between the source and the observer.
Equation (\ref{redshift}) remarkable coincides with the observed Hubble
redshift law. Therefore, if the considered generalization of the
relativistic theory is indeed realized in our Universe, the redshift in
radiation of distant objects can be, completely or in part,  
of a non-Doppler nature. It was this analysis of the redshift that 
earlier motivated us to consider $\kappa$ positive.

In the present paper we consider the properties of wave and quantum
processes within the framework of the projective theory of relativity. 
In Sec. 2, the basic relations of the theory are formulated. 
In Sec. 3, the Lagrangian equations of motion are derived. 
The next section deals with propagation of electromagnetic waves, and
the transformation rules for frequency and wave vector.
Elements of quantum mechanics are considered in the projective
relativity context in Sec. 5.
The sixth section contains the derivation of the redshift in radiation
from remote sources. 
Finally, in the last section we elaborate on the effect of contraction
of the atomic space and time scales. 

\newpage

\section{The Projective Theory of Relativity}

To obtain the space and time transformations between two inertial observers
(the Galilean transformations) one needs a certain subset of the classical 
mechanical set of axioms. If one leaves out the one that
renders the time absolute, one obtains the generalization of the Galilean 
transformations - the Lorentz transformations and the fundamental velocity
constant, $c$. If one goes further and drops the speed absoluteness axiom,
(which requires that {\it if the speeds of two particles are equal for an 
observer, they are equal for any other inertial observer}),
one arrives at so called projective Lorenz transformations
\cite{Manida1999}, \cite{Steps1999}:
\begin{equation}\label{PLSpace}
    \frac{x'}{1-\kappa c t'}=\frac{\gamma(x-vt)}{1- \kappa c t},~~~~
    \frac{y'}{1-\kappa c t'}=\frac{y}{1-\kappa c t},
\end{equation}

\begin{equation}\label{PLTime}
    \frac{t'}{1-\kappa c t'}=\frac{\gamma(t-vx/c^2)}{1-\kappa c t},
\end{equation}
where $v$ is the relative speed of the observers situated at the origins
of two inertial reference frames, 
$\gamma=1/\sqrt{1-v^2/c^2}$ is the Lorentz factor,
and $\kappa$ is a new fundamental constant.

If one considers two motionless observers,
one at the origin $\vec{x}=0$ of an inertial frame measuring
измеряющего distances and times ($\vec{x},t$), and the other in
{\bf the same} frame at the point $(x=R,y=0)$, 
measuring ($\vec{X},T$), the generalized translation transformations between 
them are as follows~\cite{Steps1999:2}:
\begin{equation}\label{SpaceTrun}
     X=\frac{x-R}{1-\kappa^2 R x}, \qquad
     Y=\frac{\sqrt{1-(\kappa R)^2}}{1-\kappa^2 R x}y.
\end{equation}

\begin{equation}\label{TimeTrun}
 1-\kappa c T=\frac{\sqrt{1-(\kappa R)^2}}{1-\kappa^2 R x}(1-\kappa c t).
\end{equation}
Equations (\ref{SpaceTrun}) have the same structure as
the velocity transformations in the conventional relativity theory. 
Thus, the components of $\vec{x}$ form a Beltrami set of coordinates
in the Lobachevski space.
Besides, neither time nor relative speeds are absolute for the two 
fixed observers. For instance, the velocities of a particle as
measured by each of these observers, $\vec{u}=d\vec{x}/dt$ and 
$\vec{U}=d\vec{X}/dT$ respectively, are related as:

\begin{equation}\label{SpeedTransf}
   U_X = \frac{u_x\sqrt{1-(\kappa R)^2}}
                {1-\kappa R u_x/c - \kappa^2 R (x-u_x t)},~~~~~
   U_Y = \frac{u_y+\kappa^2 R (yu_x-xu_y)}
                {1-\kappa R u_x/c - \kappa^2 R (x-u_x t)}.
\end{equation}
Only the objects with a zero speed possess the same speed from the point 
of view of both observers at rest.

The maximal velocity (\ref{speedlight}) is invariant under
the transformations (\ref{PLSpace}) - (\ref{TimeTrun}). For example, 
the same function of space and time
stands in the right- and left-hand sides of Eq.~(\ref{SpeedTransf}).

Transformations (\ref{PLSpace}) - (\ref{TimeTrun}) also leave
invariant the following metric
\begin{equation}\label{Metrika1}
   ds^2
   =\frac{1-(\kappa \vec{x})^2}{(1-\kappa c t)^4}c^2 dt^2
   -\frac{2\kappa \vec{x} d\vec{x}cdt}{(1-\kappa c t)^3}
   -\frac{d\vec{x}^2}{(1-\kappa c t)^2}
\end{equation}
and form a six-parametric group with generators $\vec{v},\vec{R}$
(see Appendix 1).

The expressions for the energy and momentum of a 
particle, moving at a speed $\vec{u}$ at a distance $\vec{x}$ from the 
observer, were suggested in Ref. \cite{Manida1999}:
\begin{equation}\label{Energy}
     E=\frac{m c^2}{\sqrt{1-\left[(1-\kappa c t){\displaystyle \vec{u} \over \displaystyle c} + \kappa \vec{x}\right]^2}}~,~~~
     \vec{p}=\frac{m\left[\vec{u}(1-\kappa c t) + c \kappa \vec{x}\right]}{\sqrt{1-\left[(1-\kappa c t){\displaystyle \vec{u} \over \displaystyle c} + \kappa \vec{x}\right]^2}}.
\end{equation}
They possess a number of interesting properties:

1. Both energy and momentum become infinite when 
the particle's speed approaches the speed of light 
$\vec{u} \to \vec{C}(t,\vec{x})$. Therefore, $\vec{C}(t,\vec{x})$ is the
maximal possible speed at a given point in space and time.

2. Despite time and space dependence, 
the energy and momentum of an uniformly moving particle
are constant, $d\vec{p}/dt=dE/dt=0$.
Any closed system (e.g. an atom) that is resting at the origin of the 
reference frame $\vec{x}=\vec{u}=0$ has constant energy (due to non-zero
mass) $E=mc^2$ and zero momentum $\vec{p}=0$.

3. The conventional relation holds between energy, momentum and mass:
\begin{equation}\label{Epm}
  \frac{E^2}{c^2} -\vec{p}~^2 = (mc)^2.
\end{equation}

4. The relation between energy, momentum and speed reads
\begin{equation}
  \frac{\vec{p}}{E}~c=\frac{\vec{u}}{c}(1-\kappa c t)+\kappa \vec{x},
\end{equation}
thus, for photons $\vec{u}=\vec{C}(t,\vec{x})$ and 
\begin{equation}\label{pcE}
  \vec{p}~c=\vec{n}~E,
\end{equation}
where $c$ is the fundamental speed constant, which is different from
the physical light velocity $\vec{C}(t,\vec{x})$. Therefore,
the photon momentum, as well as that of any massive particle, is
not in general collinear to its speed. Moreover, for a particle remote 
from the observer, the momentum is not zero even when the 
particle is motionless. 

5. One can easily see that for two observers in different frames the
energy-momentum transformations are given by the usual formulae:
\begin{eqnarray}
    p'_x &=& \gamma~(p_x-\frac{v}{c^2}E), \quad p'_y=p_y   \label{pELorenz1} \\
    E'   &=&\gamma~(E-v p_x).                              \label{pELorenz2}
\end{eqnarray}
For two rest observers in the same frame $E$ and $p$ 
are transformed as
\begin{eqnarray}
    P_x       &=& \sigma~(p_x-\frac{\kappa R}{c}\varepsilon),~~~P'_y=p_y  \label{pEShift1} \\
   E   &=& \sigma~(\varepsilon-c \kappa R p_x)                    \label{pEShift2}
\end{eqnarray}
where $(\varepsilon,\vec{p}~)$ are energy and momentum measured by the
observer at $(t,\vec{x})$, and $(E,\vec{P})$ - by the one 
at~$(T,\vec{X})$, and $\sigma=1/\sqrt{1-(\kappa R)^2}$.
We would like to stress the symmetry of 
transformations (\ref{pELorenz1}),(\ref{pELorenz2})
and (\ref{pEShift1}),(\ref{pEShift2}) under the substitution
$v/c \to \kappa R$.

We see that the transformations (\ref{pELorenz1})-(\ref{pEShift2}) are linear,
and this ensures that if the energy and momentum of a non-interacting particle
system are conserved in some reference frame, they will be conserved for
any other inertial observer too.

\newpage

\section{The Lagrangian Formulation}

The requirement that Lagrangian density should be invariant under the
transformations of the model leads to the following expression for $L$:
\begin{equation}\label{Lagrangian}
L(\vec{u},\vec{x},t)=-mc \frac{ds}{dt}
    =-\frac{mc^2}{(1-\kappa c t)^2} \sqrt{1-\left[\frac{\vec{u}}{c}(1-\kappa c t)+\kappa \vec{x}\right]^2}
\end{equation}

Since the Lagrangian is coordinate dependent, and, so, not
invariant under translations $\vec{x}\to\vec{x}+\vec{a}$, 
one can see that a conveniently defined momentum is not an
integral of motion either. However, the Lagrangian is invariant under
the transformations:
\begin{equation}
\vec{x}\to\vec{x}~'=\vec{x}+(1-\kappa c t)~\vec{a},
\end{equation}
where $\vec{a}$ is an arbitrary constant vector. Thus, an infinitesimal
variation of the Lagrangian
with respect to $\vec{a}$ produces a conserved quantity, which we will
call momentum:
\begin{equation}\label{LagrangianP}
\vec{p} = (1-\kappa c t) ~ \frac{\partial L}{\partial \vec{u}}
= \frac{m\left[\vec{u}(1-\kappa c t) + c \kappa \vec{x}\right]}{\sqrt{1-\left[(1-\kappa c t){\displaystyle \vec{u} \over \displaystyle c} + \kappa \vec{x}\right]^2}}.
\end{equation}

Energy conservation is related to time-invariance; the projective 
theory of relativity is invariant under the transformations
of the form:
\begin{equation}\label{shift_t1}
  \frac{t'}{1-\kappa c t'} = \frac{t}{1-\kappa c t} + t_0, ~~~
  \frac{x'}{1-\kappa c t'} = \frac{x}{1-\kappa c t},
\end{equation}
where $t_0$ is an arbitrary time shift. This invariance leads to the
conserved energy:
\begin{equation}
   E=\vec{p} \left[\vec{u}(1-\kappa c t) + \kappa c \vec{x}\right] - (1-\kappa c t)^2 L.
\end{equation}
Upon substitution of the Lagrangian density into this equation, the
expression for the energy~(\ref{Energy}) is exactly reproduced.

Let us note that transformations (\ref{shift_t1}) 
also leave unchanged the following modified Lagrangian:
\begin{equation}
       L=L_0-\frac{V(x/(1-\kappa c t))}{(1-\kappa c t)^2},
\end{equation}
where $L_0$ is the free Lagrangian. Thus, the energy of the particle
in the external field is conserved, provided that:
\begin{equation}\label{EnergyV}
  E=\sqrt{m^2 c^4 + c^2 \vec{p}^2}+V\left(\frac{\vec{x}}{1-\kappa c t}\right).
\end{equation}
We will show below that a point charge, for instance, creates field potential
of this kind.

Let us finally note that in the central field there is another 
integral of motion
\begin{equation}\label{Moment}
  \vec{M} = \frac{\vec{x}\times \vec{p}}{1-\kappa c t},
\end{equation}
which is the projective relativistic generalization of the angular momentum.

\section{Electromagnetic Wave Propagation}

Let us consider a simple thought experiment. The observer
situated at the frame origin $x=0$ sends out a series of equally separated 
(by time intervals $\tau$) light pulses along the $x$ axis.
According to~(\ref{speedlight}), the speed of these signals increases: 
$C(t,0)=c/(1-\kappa c t)$.
The trajectories of two impulses emitted at the moments
$t_0$ and $t_0+\tau$ are, respectively:
\begin{eqnarray}
  x_1(t)&=&C(t_0,0) (t-t_0)               \\
  x_2(t)&=&C(t_0+\tau,0) (t-t_0-\tau)
\end{eqnarray}
Consider now a fixed point $x$ and measure the time $\Delta t$ between
these two signals arrive at it. One can see that the frequency
$\nu_t=1/\Delta t$ grows with time:
\begin{equation}\label{nu_t}
   (1-\kappa c t) \nu_t =   (1-\kappa c t_0)\nu_0,
\end{equation}
where $\nu_0=1/\tau$.

The corresponding ``wavelength'', that is, the distance between two consequent
impulses at the moment $t$, is:
\begin{equation}
  \lambda_t =  x_2(t) - x_1(t) = \frac{C(t_0+\tau,0)}{\nu_t}.
\end{equation}
Taking the limit $\tau\to 0$ and recalling that the light speed is
constant along the trajectory of the signal $C(t_0,0)=C(t,\vec{x})$, we
obtain the relation between frequency and wavelength:
\begin{equation}
  \lambda(t,x)\nu(t,x) = C(t,x).
\end{equation}
Therefore, in the projective theory of relativity not only 
the light speed is time-space dependent, so are both the frequency
and wavelength of propagating light.

\vskip 0.5cm

Examine now a more general case of electromagnetic wave propagation.
The velocity of light $\vec{C}(t,\vec{x})$ is a function of space and time.
It is naturally to assume that the field strength in the traveling wave is 
periodic in both time and space, and its phase is given by
\begin{equation}\label{phi}
  \phi(t,\vec{x}) =\vec{a} ~ (\vec{x}-\vec{C}(t,\vec{x})t) = \vec{a} ~ \frac{\vec{x}-c\vec{n}t}{1-\kappa c t},
\end{equation}
where $\vec{a}=a\vec{n}$ is an arbitrary constant vector,
connected to the frequency and wave vector. In other words, we assume that 
$\vec{C}(t,\vec{x})$ is the phase velocity of the wave.

We proceed to determine how the phase (\ref{phi}) is affected by the transformations
(\ref{PLSpace}) - (\ref{TimeTrun}). It turns out that,
under the projective Lorentz transformations (\ref{PLSpace})-(\ref{PLTime}),
the phase will only then be invariant
\begin{equation}
\phi(t',\vec{x}')= \phi(t,\vec{x}),
\end{equation}
if the quantities $\vec{n}$ and $a$ obey the following
transformation laws:
\begin{eqnarray}\label{naTransf1}
  \vec{n}~' &=& \frac{1}{\gamma}~\frac{\vec{n}+(\gamma-1)\vec{v}(\vec{n}\vec{v})/v^2  -\gamma\vec{v}/c }{1-\vec{n}\vec{v}/c},
  \\
  a~' &=& \frac{1-\vec{v}\vec{n}/c}{\sqrt{1-v^2/c^2}} ~ a,
  \label{naTransf2}
\end{eqnarray}
where we have used the vector form of 
the projective Lorenz transformations, given in Appendix 1.

An analogous calculation shows that the wave phase is shifted by a 
constant under the projective translations:
\begin{equation}
\phi(T,\vec{X})= \phi(t,\vec{x})+\phi_0
\end{equation}
if the following conditions are satisfied:
\begin{eqnarray}\label{NATransf1}
  \vec{N} &=& \frac{1}{\sigma}~\frac{\vec{n}+(\sigma-1)\vec{R}(\vec{n}\vec{R})/R^2 -\sigma\kappa \vec{R} }
                   {1-\kappa \vec{R}\vec{n}}
  \\
  A &=& \frac{1-\kappa\vec{R}\vec{n} } 
                     {\sqrt{1-(\kappa R)^2}} ~ a,
\label{NATransf2}
\end{eqnarray}
where $\sigma=1/\sqrt{1-(\kappa R)^2}$.
Note once again the symmetry of Eqs.~(\ref{naTransf1},\ref{naTransf2}) and
(\ref{NATransf1},\ref{NATransf2}) revealed by the substitution 
$\vec{v}/c\to\kappa\vec{R}$. 
The relation (\ref{NATransf1}) provides the vector form for the light
aberration formulae obtained in Ref. \cite{Steps1999:2}.

Evidently, the constant vector $\vec{a}$ is related to the wave vector and
frequency of the propagating electromagnetic wave. 
Since the phase $\phi(t,\vec{x})$ depends on the space-time coordinates
in a non-linear fashion, we define the wave vector $\vec{k}$ and
frequency $\omega$ as follows:
\begin{eqnarray}\label{kat}
  \vec{k} &=& \frac{\partial\phi(t,\vec{x})}{\partial \vec{x}}=\frac{\vec{a}}{1-\kappa c t},
  \\
  \omega  &=&-\frac{\partial\phi(t,\vec{x})}{\partial t}=ac~\frac{1-\kappa \vec{x}\vec{n}}{(1-\kappa c t)^2}.
  \label{omat}
\end{eqnarray}
They are interconnected by the relation:
\begin{equation}\label{disper}
  \omega(t,\vec{x})=\vec{k}(t,\vec{x})\vec{C}(t,\vec{x}).
\end{equation}

It follows from Eq. (\ref{omat}) that 
$\omega = a \vec{n}\vec{C}(t,\vec{x})/(1-\kappa c t)$
and, taking into account the light speed constancy along the
trajectory, we are again led to the conclusion that the speed of light 
increases with time~(\ref{nu_t}). 

Using the equations (\ref{naTransf1})-(\ref{NATransf2}) 
one can obtain the transformation rules for frequency and wave vector.
For two observers at rest in the same reference frame, 
space-time translations lead to:
\begin{eqnarray}\label{Omegaomega}
  \Omega  &=&   \frac{1-\kappa^2 \vec{x}\vec{R}}{\sqrt{1-(\kappa R)^2}}~ \omega,
  \\
  (1-\kappa c T)\vec{K} &=& (1-\kappa c t ) \left[ \vec{k}
                           + (\sigma-1)\frac{\vec{R}(\vec{k}\vec{R})}{R^2}
                           - \sigma \kappa\vec{R} (\vec{n}\vec{k})\right],   \label{Kk}
\end{eqnarray}
where ($\omega,~\vec{k}$) are the frequency and the wave vector measured by
the observer at the origin, and ($\Omega,~\vec{K}$) - those measured by
the one at $\vec{x}=\vec{R}$.
It is possible to derive the transformation law for the frequency
directly from the coordinate transformations, 
by setting $\omega=2\pi/\Delta t$ 
and assuming that both $\vec{x}$ and $\vec{X}$ are unchanged. 

The properties of $\omega$ and $\vec{k}$ under the projective 
Lorentz transformations can be obtained from
Eqs.~(\ref{Omegaomega},\ref{Kk}) simply by the substitution $\kappa \vec{R} 
\to \vec{v}/c$.

\section{The Quantum Theory}

Constructing a consistent quantum theory requires a careful
consideration of its axiomatic foundation and of the amendments
that are necessary within the context of projective
relativity. Here, we undertake a more modest task--- 
to consider some heuristic conjectures, which will allow us
to obtain quantum mechanical relations that could be expected 
to hold in the complete theory too.

Let us find what changes are required to the Planck---Einstein 
relations for the light quanta. To this end, we will proceed from the 
requirement of relativistic
invariance, in analogy with the well-known argumentation by de Broglie.
We write the transformations for the momentum (\ref{pEShift1}) 
in the standard vector form (Appendix 1) and taking into account that 
Eq.~(\ref{pcE}) gives $\varepsilon=c(\vec{p}\vec{n})$ when applied to
light, we obtain the 
following relation between the values of the photon momentum, as
measured by two observers in the same inertial reference frame:

\begin{equation}
   \vec{P}= \vec{p} + (\sigma-1)
                    \frac{\vec{R}(\vec{p} \vec{R} )}{R^2}
                    - \sigma\kappa\vec{R}(\vec{n}\vec{p}).
\end{equation}
Comparing this expression with that for the wave vector (\ref{Kk}),
we can conclude that the simplest covariant connection between them
can be written as
\begin{equation}\label{phbar}
   \vec{p}= (1-\kappa c t)\hbar \vec{k} = \hbar \vec{a}.
\end{equation}
Note that the momentum of the quantum particle is constant, as it is for the
classical one.

Substituting the wave vector (\ref{phbar}) expressed in terms of
the particle momentum into Eq.~(\ref{disper}) and using~(\ref{pcE}) 
we obtain a generalized Plank formula:
\begin{equation}\label{Plank}
   E= \frac{(1-\kappa c t)^2}{1-\kappa \vec{x}\vec{n}} \hbar \omega = \hbar a.
\end{equation}
The relations (\ref{phbar}),(\ref{Plank}) are ``cosmological''
generalizations of the standard quantum-mech\-an\-ical expressions
$E=\hbar \omega$, $\vec{p}=\hbar \vec{k}$.

One could also derive these relations within the framework of conventional
canonical quantization. Indeed, it follows from Eq.~(\ref{LagrangianP}), 
that the canonical momentum is related to the wave
vector in a usual way:
\begin{equation}
   \vec{\pi} = \frac{\partial L}{\partial \vec{u}} = \frac{\vec{p}}{1-\kappa c t} = \hbar \vec{k}.
\end{equation}
Besides, the Hamiltonian (which, unlike the energy, is not conserved!) 
is conveniently connected to the frequency of light emitted 
during a quantum transition:
\begin{equation}
   H = \vec{\pi}\vec{u} - L = \frac{E - \kappa c \vec{x}\vec{p} }{(1-\kappa c t)^2} = \hbar\omega.
\end{equation}

One can speculate that the standard commutation rule holds 
for the canonical momentum operator:
\begin{equation}
     [\pi_\alpha, x_\beta] = \frac{\hbar}{i}~\delta_{\alpha\beta},
\end{equation}
and, thus, the momentum operator $p$ introduced above has coordinate 
representation
\begin{equation}
    \vec{p} = (1-\kappa c t)\frac{\hbar}{i}\vec{\nabla}.
\end{equation}
In this case, the phase of the free particle wave function is
analogous to the phase of the electromagnetic wave
\begin{equation}
    \psi_p \sim \exp\frac{i}{\hbar}~\frac{\vec{p}\vec{x}-Et}{1-\kappa c t}.
\end{equation}
In Sect. 7 below, we will consider the quantization of a hydrogen atom,
and some consequences of space-time scale contraction, which are relevant 
to this problem.

\section{The Redshift}

The most spectacular result of the projective theory of relativity
is the fact that the frequency of the radiation changes, and this
frequency shift increases with the distance
from the source. We will rederive this result here,
slightly changing the approach pursued in Ref. \cite{Steps1999:2}.

Let us place the Earth laboratory at the frame origin $\vec{x}=0$, and the
remote source at a distance $R$ from it ($\vec{x}=\vec{R}$). 
At the moment when the light signal is emitted by the source, the frequency
measured by local and remote observers, $\omega_1$ and $\Omega_1$, are 
related, according to Eq.~(\ref{Omegaomega}), by the expression:
\begin{equation}
   \Omega_1 = \sqrt{1-(\kappa R)^2} \omega_1.
\end{equation}
The frequency of the propagating electromagnetic wave increases (\ref{nu_t}) 
and, when the signal reaches the laboratory observer at time $t_2$,
becomes $\omega_2$:
\begin{equation}
   (1-\kappa c t_2)\omega_2 = (1-\kappa c t_1)\omega_1.
\end{equation}

The relation between the time $t_1$ when the signal was emitted (according to 
our watch) and the time $t_2$ when it arrived can be easily found if one
recalls that the light speed is constant along the trajectory
$C(t_1,R)=C(t_2,0)$, $\vec{n}= - \vec{R}/R$:
\begin{equation}\label{t1_t2}
   1-\kappa c t_1 = (1+\kappa R)(1-\kappa c t_2).
\end{equation}

Now we can obtain the expression linking the frequency $\Omega_1$ measured by
the remote observer and the frequency $\omega_2$ of the signal that reaches 
Earth:
\begin{equation}\label{Habble01}
   \frac{\Omega_1}{\omega_2} = \sqrt{\frac{1-\kappa R}{1+\kappa R}}.
\end{equation}
It was the intention to reproduce the observed Hubble effect that motivated
the choice of the negative sign for $\kappa$ in 
Refs.~\cite{Manida1999}-\cite{Steps1999:2}.

This derivation, however, should be further continued to incorporate the
quantum mechanical relations obtained above. When one studies the redshift
in the spectra of remote galaxies, one compares the observed frequency 
$\omega_2$ 
with the spectral frequencies $\omega_0$ of the identical atoms 
on the Earth, and not with $\Omega_1$.
We thus look at the Universe at a certain moment in the past; the
time of emission by the local watch $T_1$ is connected with the present time
on the Earth (\ref{TimeTrun}), (\ref{t1_t2}) by the formula
\begin{equation}
1-\kappa c T_1 = \frac{1-\kappa c t_1}{\sqrt{1-(\kappa R)^2}} 
               = \sqrt{\frac{1+\kappa R}{1-\kappa R}} ~(1-\kappa c t_2).
\end{equation}
Since the frequency of the radiation changes with time, we see that the frequency
$\Omega_1$ is lower then that of the same atom on Earth at present:
\begin{equation}
  \Omega_1 = \frac{E/\hbar}{(1-\kappa c T_1)^2}, \quad 
  \omega_0 = \frac{E/\hbar}{(1-\kappa c t_2)^2},
\end{equation}
where $E$ is the radiation energy of one of the two identical motionless 
atoms in the vicinity of the observer.
Therefore,
\begin{equation}
   \frac{\Omega_1}{\omega_0} = \frac{1-\kappa R}{1+\kappa R}.
\end{equation}

Substituting this relation into Eq. (\ref{Habble01}) we finally obtain the 
correct expression for the redshift in the atomic spectra:
\begin{equation}
   \frac{\omega_0}{\omega_2} = \sqrt{\frac{1+\kappa R}{1-\kappa R}}.
\end{equation}

One can see that this formula conforms with the redshift observed in
experiment ($\omega_2 > \omega_0$) if one chooses the constant $\kappa$ to be 
positive, $\kappa>0$. 

If one assumes that the cosmological redshift is completely accounted for
by the effects of the projective theory of relativity, the fundamental constant 
$\kappa$ is directly related to the Hubble constant:
\begin{equation}
   \kappa c = H = 65 ~\frac{km/sec}{Mps} = \frac{6.7~~10^{-11}}{year}.
\end{equation}

As we noted above, the velocity of light increases with time and at the moment
$t_0=1/\kappa c$ it will become infinite. This future event is the singularity 
of the considered theory. There is no matter singularity in the Universe 
described by the
projective relativity. The Universe existed infinitely long in the past;
the gradual change of the light velocity from its initial zero value
was accompanied by various evolutionary processes.

A question arises about the behaviour of matter in this Universe. Gravitational
forces can contribute to the redshift due to
the convenient cosmological model mechanism; 
this contribution can be of either 
sign. There is, however, another possibility. 
Under the influence of gravity, the matter gradually concentrates around
initial inhomogeneities, whereas remaining on average at rest at large scales. 
This process, becoming increasingly faster as the speed of light grows,
leads to the formation of stars, galaxies, etc.

\section{Expanding Universe vs Contracting Space-Time Scales}

Let us consider quantization of a hydrogen atom within the framework of
the projective theory of relativity. The electromagnetic field action
invariant under the projective transformations can be written as
\begin{equation}
   S  = -mc\int d s - \frac{e}{c} \int A_\nu dx^\nu - \frac{1}{16\pi c}\int F_{\mu\nu}F^{\mu\nu} \sqrt{-g}~d^4x.
\end{equation}
The components of the metric tensor are, from Eq.~(\ref{Metrika1})
\begin{equation}
  g_{00} = \frac{1-(\kappa \vec{x})^2}{(1-\kappa c t)^4}, ~~
  g_{0i} = -\frac{\kappa x^i}{(1-\kappa c t)^3}, ~~
  g_{ij} = -\frac{\delta_{ij}}{(1-\kappa c t)^2}
\end{equation}
and, so, the invariant four-volume is $d^4x\sqrt{-g}$ $=$ $d^4x / (1-\kappa)^5$,
and  $F_{\mu\nu}=\partial_\mu A_\nu - \partial_\nu A_\mu$.

The field equation describing a set of point charges has the usual form
\cite{Landau}:
\begin{equation}\label{FLandau}
   \partial_\alpha\left(\sqrt{-g}F^{\alpha\beta}\right) = 4\pi \sum e_a \delta(\vec{x}-\vec{x}_a)\frac{dx^\beta}{dx^0}.
\end{equation}
When the magnetic field is absent, $A_i = 0$, Eq. (\ref{FLandau})
is reduced to the Poisson type equation:
\begin{equation}
   (1-\kappa c t)\Delta A_0 = - 4\pi e_a \delta(\vec{x}).
\end{equation}
So, the contribution to the Lagrangian from the interaction between a
charged particle and the field of a point charge,
\begin{equation}
    L_{int} = \frac{e}{c} A_\nu \frac{dx^\nu}{dt} = \frac{e^2 }{(1-\kappa c t)r}
\end{equation}
in accordance with (\ref{EnergyV}) results in the conserved energy
\begin{equation}
    E = \sqrt{m^2 c^4 + c^2\vec{p}^2} - \frac{(1-\kappa c t)e^2 }{r}.
\end{equation}
One can note a formal coincidence between the obtained ``time-dependence''
of charge and Gamow's hypothesis~\cite{Gamov1967}.

The operator expression for the energy eigenvalues equation has the form:
\begin{equation}
    \left[\sqrt{m^2 c^4 - c^2\hbar^2(1-\kappa c t)^2 \Delta} - \frac{(1-\kappa c t)e^2 }{r}\right] \psi_n = E_n \psi_n.
\end{equation}
Performing the substitution $x _i \to x_i(1-\kappa c t)$ one can easily see
that energy levels are stationary and given by the standard formula
\begin{equation}
      E_n = \frac{m e^4}{\hbar^2} ~f_n\left(\frac{e^2}{\hbar c}\right),
\end{equation}
while the eigenfunctions are time-dependent. In particular,
the mean value of electron position operator in a central field decreases
with time:
\begin{equation}
       <r> = <r>_0 (1-\kappa c t).
\end{equation}
This result is in fact quite general for the closed systems with 
conserved energy (\ref{EnergyV}). 
Therefore, all length units that are based upon
properties of atomic objects tied by the electromagnetic forces,
also decrease with time \cite{Hoyle1971}.

If we take the wavelength and frequency of the atomic spectra as 
the standard length and time, we would have
\begin{equation}\label{etalon}
   \Delta\tau(t) = \frac{1}{\omega}=\frac{(1-\kappa c t)^2}{E/\hbar}, ~~~
   \Delta l(t) = \frac{\lambda}{2 \pi}=\frac{1-\kappa c t}{E/\hbar c}
\end{equation}
and the variation in the light 
velocity would be non-observable~\cite{ManidaPriv}.
A note is in order here that, since the standard second determined from
frequency shortens with time, the future singularity at $t=1/\kappa c$ 
cannot be reached if we measure time with the clock (\ref{etalon}).

On the atomic scale, one could change the notation using the coordinates 
introduced in Ref. \cite{Manida1999}:
\begin{equation}
    \tilde{t}=\frac{t}{1-\kappa c t}, \qquad \tilde{x}=\frac{x}{1-\kappa c t}.
\end{equation}
In terms of these ``atomic variables'' $(\tilde{x},\tilde{t})$,
the transformations between two inertial frames (\ref{PLSpace}), 
(\ref{PLTime}) become the conventional Lorentz transformations.
The translational transformations between two spatially separated rest 
observers
(\ref{SpaceTrun}), (\ref{TimeTrun})
can also be written in the Lorenz form:
\begin{equation}\label{SpaceTrun2}
  \tilde{X} = \frac{\tilde{x} - R - v \tilde{t} }{\sqrt{1- 
 {\displaystyle v^2 \over \displaystyle c^2}} }, \quad \tilde{Y}=\tilde{y},
\end{equation}
\begin{equation}\label{TimeTrun2}
    \tilde{T} = \frac{\tilde{t} - \tilde{t}_0 - v \tilde{x}/ c^2 }{\sqrt{1-{\displaystyle v^2 \over \displaystyle c^2}}},
\end{equation}
However, the formal velocity parameter entering these formulae is actually
the distance between the observers:
$v/c = \kappa R$, а $\tilde{t}_0=(1-\sqrt{1-(\kappa R)^2})/\kappa c$.
Thus, in terms of ($\tilde{x}, \tilde{t}$) the translational
transformations
describe two observers that are moving away from each other at a speed
proportional to the distance between them.
In contrast to the original transformations~(\ref{SpaceTrun},\ref{TimeTrun}), 
the interpretation of Eq.~(\ref{SpaceTrun2},\ref{TimeTrun2}) is not
immediately obvious. The fact that the two observers
in (\ref{SpaceTrun},\ref{TimeTrun}) have zero relative velocity is 
verified by repeated measurements and distinguish these observers from
the others. The origin of the redshift lies in the time transformation between
them. On the other hand, when expressed through variables ($\tilde{x}, 
\tilde{t}$), this effect has a Doppler nature and is intrinsic 
to Eqs.~(\ref{SpaceTrun2}),(\ref{TimeTrun2}).

One can treat the transformations (\ref{SpaceTrun2}),(\ref{TimeTrun2})
as the basis of the theory, provided that an {\bf additional} 
postulate is introduced: $v/c = \kappa R$.
Indeed, assuming that the Universe is governed by the standard theory
of relativity, save that any two observers are moving away from each
other at a speed proportional to the distance between them, we obtain
the Eq. (\ref{SpaceTrun2}),(\ref{TimeTrun2}).
Let us note that the parameter $\tilde{t}_0$ necessarily arises
in the described theory. It must be present if we require that the 
transformations compose a group. Due to its presence, all the observers 
are equal, 
and the Big Bang occurs at the same moment according to all local 
atomic clocks $\tilde{T}_{BB}=\tilde{t}_{BB}=-1/\kappa c$.
In terms of ($x, t$) this singularity is absent, because it corresponds to
the infinitely distant moment in the past.

\vskip 1cm

{\bf Conclusion}

The initial motivation for developing the projective theory of relativity 
was the desire to build a parametric generalization of special
relativity. The basic relations of the theory
(\ref{PLSpace} - \ref{TimeTrun}) 
were obtained by using only universal arguments and without introducing
any additional axioms (the principle of parametric incompleteness).
The variability of the light velocity and the cosmological redshift
arose within this model as quite unexpected consequences.
It was shown in the present paper that, under the plausible assumptions,
the sizes of atomic systems decrease with time.
 
We thus face the well known dilemma of the expanding Universe: 
either the Universe is unchanged and our standard rules are contracting,
or the units of length are constant and the Universe is expanding. 
Whichever point of view one prefers, there is a redshift in the spectra
of distant objects.

The question as to what is the ``true'' constant length, cannot be 
unambiguously answered at
present. It could be the ``radius of the Universe'' $1/\kappa$, 
and could be the cesium atomic wavelength. 
If the space-time is indeed quantized, the size of this quantum would
play this role and provide the solution to the outlined problem.

If the projective relativity is indeed realized in the world we observe,
the cosmological redshift is as fundamental as the existence of the
highest possible speed. In particular, it does not depend on the
matter density and the initial conditions in the Universe, and is an
intrinsic property of the space itself.

\vskip 0.5cm

{\bf Acknowledgement}

I would like to thank Prof. Manida for the incisive comments he made in
the communication with the author \cite{ManidaPriv}. Some of the result
reported here were also independently obtained by Prof. Manida.

\newpage
\begin{center}
{\bf Appendix 1.}\\
{\bf \it Some vector relations of the Projective STR}
\end{center}

The projective Lorentz transformations can be written in the standard
vector form:
\begin{eqnarray}\label{PrimL1}
   \frac{\vec{x}'}{1-\kappa c t'}&=&\frac{1}{1-\kappa c t}\left[
      \vec{x}+(\gamma -1)\frac{\vec{x}(\vec{x}\vec{u})}{v^2}-\gamma \vec{v} t
   \right],\\
   \frac{t'}{1-\kappa c t'}&=&\gamma\frac{ t-\vec{x}\vec{v}/c^2}{1-\kappa c t}
   \label{PrimL2}
\end{eqnarray}

The projective translational transformations between two observers at
rest are:
\begin{eqnarray}\label{PrimT1}
   \vec{X}&=&\frac{\vec{x}-\vec{R}+\left(1-\sqrt{1-(\kappa R)^2}\right)~\vec{R}\times(\vec{R}\times\vec{x})/R^2}{1-\kappa^2 \vec{x}\vec{R}},
   \\
   1-\kappa c T&=&\frac{\sqrt{1-(\kappa R)^2}}{1-\kappa^2 \vec{x}\vec{R}}~(1-\kappa c t).
   \label{PrimT2}
\end{eqnarray}

Under the projective transformations, Manida's 4-vector
\begin{equation}
d\tilde{x}^\mu = d \left(\frac{ct}{1-\kappa c t},~ \frac{\vec{x}}{1-\kappa c t}\right)
               = \left(\frac{cdt}{(1-\kappa c t)^2},~
                 \frac{d\vec{x}(1-\kappa c t)+\kappa c \vec{x}dt}{(1-\kappa c t)^2} \right)
\end{equation}
is transformed by the usual linear Lorenz rules. For the rest observers
the transformations have the form:
\begin{eqnarray}
d\tilde{X}^0 &=& \sigma\left(d\tilde{x^0}-\kappa \vec{R}d\tilde{\vec{x}}\right),\\
d\tilde{\vec{X}}&=& d \tilde{\vec{x}} + \left(\sigma-1\right)
                   \frac{\vec{R}(\vec{R} d\tilde{\vec{x}})}{R^2}
                 - \sigma \kappa \vec{R}d\tilde{x}^0,
\end{eqnarray}
where $\sigma=1/\sqrt{1-(\kappa R)^2}$.

The transformational rules for the energy and momentum $p^\nu=mc
d\tilde{\vec{x}}/ds = (\varepsilon, \vec{p})$ under space-time translations
can be written as:
\begin{eqnarray}\label{PrimP1}
E &=& \sigma\left(\varepsilon- c \kappa \vec{R}\vec{p}\right),\\
\vec{P}&=& \vec{p} + (\sigma-1)
                   \frac{\vec{R}(\vec{R} \vec{p})}{R^2}
                 - \sigma\kappa \vec{R}~\frac{\varepsilon}{c}.\label{PrimP2}
\end{eqnarray}
One can derive the Lorenz rules for $(\varepsilon, \vec{p})$ from the Eqs.
(\ref{PrimP1}), (\ref{PrimP2}) by substituting $\kappa \vec{R} \to \vec{v}/c$.

Let us note also the following invariant of translational transformations 
(\ref{PrimT1}),(\ref{PrimT2}):
\begin{equation}
\frac{1-\kappa c T}{\sqrt{1-(\kappa \vec{X})^2}} = \frac{1-\kappa c t}{\sqrt{1-(\kappa \vec{x})^2}} = inv.
\end{equation}


\newpage

\vskip 1cm
\begin{thebibliography}{50}

\bibitem{Einstein1905}
   {\it Einstein A.} -
   "Zur Elektrodynamik bewegter Korper"
   Ann.d.Phys., b.17, s.891, {\bf 1905.}
%%%%%%%%%%%%%%%%%%%%%%   Relativity without light 1. %%%%%%%%%%%%%%%%%%%%%%%%%
\bibitem{Pauli}
   {\it Ignatowsky W.} -
   Arch.Math.Phys., Vol. 17, {\bf 1910.} p.1; Vol. 18 {\bf 1911.} p.17;
   {\it Frank P.  Rothe H.} -
   Ann.Phys., Vol. 34, {\bf 1911.} p.825.


%%%%%%%%%%%%%%%%%      Projectiv Realtive Theory      %%%%%%%%%%%%%%%%%%%%%%%
\bibitem{Manida1999}
  {\it Manida S.N.}
  "Fock-Lorentz transformations and time-varying speed of light"
  gr-qc/9905046  {\bf 1999}

\bibitem{Steps1999}
  {\it Stepanov S.S.}
  "Fundamental Physical Constants and the Principle of Parametric Incompleteness"
  physics/9909009  {\bf 1999}

\bibitem{Steps1999:2}
  {\it Stepanov S.S.}
  "A time-space varying speed of light and the Hubble Law in static Universe"
  Phys.Rev.D, v.61, p.1235XX,{\bf 2000} (astro-ph/9909311)

\bibitem{ManidaPriv}
  {\it Manida S.N.}
  "Private correspondence",  {\bf 1999}

%%%%%%%%%%%%%%%%%%%         Time-varying speed of light    %%%%%%%%%%%%%%%%%%

\bibitem{Moffat}
   {\it Moffat J.W., } -
   "Superluminary Universe: A Possible Solution to the Initial Value
   Problem in Cosmology"
   Int.J.Mod.Phys., Vol. 2D, No.3 {\bf 1993.} p. 351-365
   gr-qc/9211020 {\bf 1992.}

\bibitem{Albrecht1}
   {\it Albrecht A., Magueijo J.} -
   "A time varying speed of light as a solution to cosmological puzzles"
   Phys.Rev., Vol. 59D, {\bf 1999.} p. 43516,
   astro-ph/9811018 {\bf 1998.}


\bibitem{Barrow1}
   {\it Barrow J.D.} -
   Phys.Rev. D 59, 000 {\bf 1999};
   {\it Barrow J.D., Magueijo J.} -
   "Varying -$\alpha$ Theories and Solutions to the Cosmological Problems"
   astro-ph/9811072 {\bf 1998.}


\bibitem{Barrow2}
   {\it Barrow J.D., Magueijo J.} -
   "Solving the Flatness and Quasi-flatness Problems in Brans-Dicke
   Cosmologies with a Varying Light Speed"
   astro-ph/9901049 {\bf 1999.}

\bibitem{Barrow3}
   {\it Barrow J.D., O'Toole C.} -
   "Spatial Variations of Fundamental Constants"
   astro-ph/9904116 {\bf 1999.}

\bibitem{Avel99}
   P. P. Avelino and  C. J. A. P. Martins,
   Phys.Lett. {\bf B459}, 468 (1999).


\bibitem{Albrecht2}
   {\it Albrecht A. } -
   "Cosmology with a time-varying speed of light"
   astro-ph/9904185 {\bf 1999.}

\bibitem{Brandenberger1999}
   {\it Brandenberger R.H., Magueijo J.} -
   "Imaginative Cosmology"
    hep-ph/9912247 {\bf 1999.}


\bibitem{Youm}
   {\it Donam Youm} -
   "Brane world cosmologies with varying speed of light"
   Phys.Rev. D 63, 125011 {\bf 2001};

\bibitem{Landau}
   {\it L. D. Landau, E. M. Lifshitz} - 
   "The Classical Theory of Fields"
    402 p. 4th Edition Vol 2 {\bf 1997}.



\bibitem{Gamov1967}
   {\it Gamov G.} -
   "Electricity, gravity and cosmology"
    Phys.Rev.Lett 19, 759 {\bf 1967}.

\bibitem{Hoyle1971}
   {\it Hoyle F., Narlikar J.W.} -
   "On the nature of mass"
    Nature 233, 41 {\bf 1971}.


\end{thebibliography}
\end{document}